\documentclass[twocolumn,showpacs,floatfix,superscriptaddress,amsmath,amssymb,pre]{revtex4}
\usepackage{mathrsfs}
\usepackage{txfonts}
\usepackage{amssymb}
\usepackage{graphicx}
\usepackage{hyperref}
\usepackage{ulem}
 \usepackage{overpic}
 \usepackage{psfrag}
\usepackage{tabularx}
\usepackage{array}
\usepackage{placeins}
\makeatletter
\renewcommand\subsection{\@startsection
{subsection}{1}{0mm}
 {-\baselineskip}
 {0.5\baselineskip}
{\FloatBarrier\normalfont\Large\bfseries}}
\makeatother
\newcommand{\be}{\begin{equation}}
\newcommand{\ee}{\end{equation}}
\newcommand{\PreserveBackslash}[1]{\let\temp=\\#1\let\\=\temp}

\begin{document}
\title{Degenerate groundstates and multiple bifurcations
       in a two-dimensional $q$-state quantum Potts model}

\author{Yan-Wei Dai}
\affiliation{Centre for Modern Physics and Department of Physics,
Chongqing University, Chongqing 400044, The People's Republic of
China}

\author{Sam Young Cho}
\email{sycho@cqu.edu.cn}
\affiliation{Centre for Modern Physics and Department of Physics,
Chongqing University, Chongqing 400044, The People's Republic of
China}

\author{Murray T. Batchelor}
\affiliation{Centre for Modern Physics and Department of Physics,
Chongqing University, Chongqing 400044, The People's Republic of
China}
\affiliation{Australian National University, Canberra ACT 0200, Australia}

\author{Huan-Qiang Zhou}
\affiliation{Centre for Modern Physics and Department of Physics,
Chongqing University, Chongqing 400044, The People's Republic of
China}

\begin{abstract}
 We numerically investigate the two-dimensional $q$-state quantum Potts model
 on the infinite square lattice by using
 the infinite projected entangled-pair state (iPEPS) algorithm.
 We show that the quantum fidelity, defined as
 an overlap measurement between
 an arbitrary reference state and the iPEPS groundstate of the system,
 can detect $q$-fold degenerate groundstates for the $Z_q$ broken-symmetry phase.
 Accordingly, a multiple-bifurcation of the quantum groundstate fidelity is shown to occur as
 the transverse magnetic field varies from the symmetry phase
 to the broken-symmetry phase, which means that a multiple-bifurcation point
 corresponds to a critical point.
 A (dis-)continuous behavior of quantum fidelity
 at phase transition points characterizes a (dis-)continuous phase
 transition.
 Similar to the characteristic behavior of the quantum fidelity,
 the magnetizations, as order parameters, obtained from the degenerate groundstates
 exhibit multiple bifurcation at critical points.
 Each order parameter is also explicitly demonstrated to transform under
 the subgroup of the $Z_q$ symmetry group.
 We find that the $q$-state quantum Potts model on the square lattice
 undergoes a discontinuous (first-order) phase transition for $q=3$ and $q=4$, and
 a continuous phase transition for $q=2$ (the 2D quantum transverse Ising model).

\end{abstract}

\pacs{05.30.Rt, 03.67.-a, 05.50.+q, 75.40.Cx}
 %

\maketitle

\section{Introduction}
 Most quantum phase transitions \cite{sachdev} in quantum many-body physics
 can be understood within the Landau-Ginzburg-Wilson paradigm which
 provides the fundamental key concepts of
 spontaneous symmetry-breaking and local order parameters.
 In the last decades, some of the most remarkable discoveries,
 such as various magnetic orderings,
 the integer and fractional quantum Hall effects \cite{hall1,hall2}
 and high-$T_c$ superconductors \cite{superconductor}
 have brought more attention to quantum phase transitions
 in condensed matter physics.
 However, some systems do not seem to be well understood within the paradigm
 in characterizing newly discovered quantum states.
 Also, in spite of the decisive role of the key concepts in characterizing quantum phase transitions,
 practical and systematic ways to
 understand some (either explicit or implicit) broken-symmetry phases and
 (either local or nonlocal) order parameters have not been readily available.
 The crucial difficulties reside in the fact that
 (i) calculating groundstate wavefunctions and identifying
     degenerate groundstate wavefunctions are usually a formidable task,
 and (ii) an efficient way to determine groundstate phase diagrams
 is necessary.

 Encouragingly, in the past few years,
 significant advances have
 been made, both in classically simulating quantum
 lattice systems and in determining groundstate
 phase diagrams \cite{iTEBD,iPEPS,jonas,PHYLi}.
 Especially, tensor network representations
 provide efficient quantum many-body wave functions
 to classically simulate quantum many-body
 systems \cite{iTEBD,iPEPS,TEBD,MERA,PEPS,MPS}.
 Tensor network algorithms in quantum lattice systems have made
 it possible to investigate their groundstates
 with an imaginary time evolution \cite{iTEBD}.
 By using two novel approaches proposed from
 a quantum information perspective --
 entanglement \cite{Entanglement1, Entanglement2, Entanglement3, Entanglement4}
  and fidelity \cite{zhou1, zhou2, Zanardi} --
 tensor network groundstates have been successfully implemented to
 determine groundstate phase diagrams
 of quantum lattice systems without
 prior knowledge of order parameters.

 Although these latest advances in understanding quantum phase transitions
 have been achieved, directly understanding degenerate
 groundstates originating from spontaneous symmetry breaking
 and connections between a symmetry breaking and its corresponding order parameter,
 as the heart of the Landau-Ginzburg-Wilson theory, still remains
 largely unexplored.
 With a randomly chosen initial state subject to an imaginary
 time evolution, the tensor network algorithms
 can offer an efficient way to
 directly investigate degenerate groundstates
 in quantum lattice systems.
 For one-dimensional spin lattice systems,
 doubly degenerate ground states for broken-symmetry phases
 have been detected by means
 of the quantum fidelity bifurcations with the tensor network algorithm
 in various spin lattice models such as the quantum Ising model,
 spin-1/2 XYX model with transverse magnetic field, among others \cite{zhouzhlb,zhwlzh,dhzhzhou}.
 Very recently, Su {\it et al.} \cite{Su}
 have further demonstrated that the quantum fidelity measured by
 an arbitrary reference state
 can detect and identify explicitly all degenerate ground states
 ($N$-fold degenerate groundstates) due to
 spontaneous symmetry breaking in broken-symmetry phases
 for the infinite matrix product state (iMPS) representation
 in the one-dimensional (1D) $q$-state quantum Potts model.
 It has been also discussed how each order parameter calculated from degenerate ground states
 transforms under a subgroup of a symmetry group of the Hamiltonian.

 In contrast to 1D quantum systems,
 however, two-dimensional (2D) quantum systems have not yet been explored to
 detect their degenerate ground states for broken-symmetry phases.
 We will thus explore the spontaneous symmetry breaking mechanism in a 2D quantum system.
 To describe a 2D many-body wavefunction,
 we will employ the infinite projected entangled-pair state (iPEPS) \cite{iPEPS,iTEBDorus,xiang}.
 The infinite time-evolving block decimation (iTEBD)
 method \cite{iTEBD} will be used to calculate iPEPS groundstate wavefunctions with
 randomly chosen 2D initial states.
 In order to distinguish degenerate groundstate wavefunctions of broken-symmetry phases
 and to determine phase transition points,
 the quantum fidelity \cite{Su},
 defined as an overlap measurement between an arbitrary 2D reference state
 and the iPEPS groundstates of the system, will be employed.
 The defined quantum fidelity
 corresponds to a projection of each 2D iPEPS groundstate
 onto a chosen 2D reference state. Consequently,
 the number of different projection
 magnitudes denotes the groundstate degeneracy of a system
 for a fixed system parameter.
 Also, a critical
 point can be noticed by the collapse of different projection magnitudes to one
 projection magnitude.
 With such a property of the quantum
 fidelity, the different projection magnitudes of the ground
 states starting from the collapse point can be called
 a multiple bifurcation of the quantum fidelity.
 Furthermore, an analysis of a relation between local observables,
 as order parameters,
 from each of degenerate groundstates can allow us to specify
 exactly which symmetry of the system is broken in the broken-symmetry phase.

 In this paper, we consider
 the 2D $q$-state quantum Potts model on the infinite square lattice with a transverse
 magnetic field.
 In general, $q$-state Potts models have been shown to exhibit fundamental
 universality classes of critical behavior and have thus
 become an important testing platform for different numerical
 approaches in studying critical phenomena \cite{potts,wu}.
 It is well known that the 2D classical Potts model and its equivalent 1D quantum Potts chain
 are exactly solved models at the critical point \cite{baxter,wu,hamer}.
 In contrast to the 1D quantum Potts model,
 the 2D quantum $q$-state Potts model on the square lattice has not been so well understood.
 However, for $q=2$ the 2D quantum transverse Ising model
 and the equivalent 3D classical Ising model
 have been widely studied via a number of different techniques
 (see, e.g., Refs.~\cite{blote,hamer2,ctmrg} and references therein).
 For $q=3$, there appears to be only one investigation \cite{hamer_q=3} of the 2D quantum Potts model,
 with however, many studies of the 3D classical version of the 3-state Potts model,
 by Monte Carlo, series expansions etc. (see, e.g.,
 Refs.~\cite{wu,herrmann,nienhuis,blote,hamer_q=3,hamer2,ctmrg} and references therein).
 As far as we are aware, there have been no studies of the 2D quantum $4$-state Potts model.

 The classical mean-field solutions~\cite{wu} and the extensive
 computations (see, e.g.,
 Refs.~\cite{wu,herrmann,nienhuis,blote,hamer_q=3,hamer2,ctmrg} and references therein)
 have suggested that the 3D classical $q$-state Potts model, and thus the
 2D quantum $q$-state Potts model,
 undergo a continuous phase transition
 for $q \leq 2$ and a first-order phase
 transition for $q>2$.
 In this paper, for the 2D quantum $q$-state Potts model on the square lattice,
 from the iPEPS groundstates calculated for fixed system parameters,
 each of the $q$-fold degenerate groundstates
 due to the broken $Z_q$ symmetry are distinguished by means of the quantum fidelity
 with $q$ branches in the broken-symmetry phase.
 A continuous (discontinuous) property of the quantum fidelity function across the phase transition
 point reveals a continuous (discontinuous)
 quantum phase transition for $q = 2$ ($q = 3$ and $q=4$).
 The multiple bifurcation points are shown to correspond
 to the critical points.
 Also, we discuss a multiple bifurcation of
 local order parameters and its characteristic properties for the
 broken-symmetry phase.
 We demonstrate clearly how the order parameters
 from each of the degenerate ground states transform under the
 subgroup of the symmetry group $Z_q$.

 This paper is organized as follows. In Sec. II, the 2D $q$-state quantum Potts model
 on the square lattice is defined.
 In Sec. III, we briefly explain the
 iPEPS representation and the iTEBD method in 2D square lattice systems.
 Section IV presents how to detect degenerate groundstates
 by using the quantum fidelity between the degenerate groundstates
 and a reference state.
 In Sec. V, quantum phase transitions are discussed based on multiple
 bifurcations and multiple bifurcation points of the quantum fidelity.
 In Sec. VI, we discuss the magnetizations given from the degenerate groundstates
 and demonstrate their relation with respect to the subgroup of
 the $Z_q$ symmetry group of the 2D $q$-state quantum Potts model on
 the square lattice.
 Our summary and concluding remarks are given in Sec. VII.

\section{Two-dimensional quantum $q$-state Potts model}

 To demonstrate detecting degenerate groundstates
 in 2D quantum lattice systems, we consider the $q$-state quantum Potts
 model~\cite{solyom} on an
 infinite square lattice in a transverse magnetic field:
\begin{equation}
  H_q =-\sum_{(\vec r, \vec r')}
  \left(\sum_{p=1}^{q-1}M^{[\vec r]}_{x\;,\;p}M^{[\vec r']}_{x\;,\;q-p}\right)
    -\sum_{\vec r} \lambda  M^{[\vec r]}_{z},
    \label{ham1}
\end{equation}
 where $\lambda$ is the transverse magnetic field and
 $M^{[\vec r]}_{\alpha,p}$ with $p \in [1, q-1]$
 ($\alpha=x,z$) are the $q$-state Potts `spins' at site $\vec r$.
 The $q$-state Potts spin matrices are given by
\begin{displaymath}
  M_{x,1}= \left(\begin{array}{cc}
                    0 & I_{q-1} \\
                    1 & 0
            \end{array} \right)
 \mbox{~and~}
  M_{z}= \left(\begin{array}{cc}
                     q-1 & 0 \\
                        0 & -I_{q-1}
            \end{array} \right),
\end{displaymath}
where  $I_{q-1}$ is the $(q-1)\times (q-1)$ identity  matrix
  and $M_{x,p} = (M_{x,1})^p$.
  $(M_{x,1})^{q}$ equals the $q\times q$ identity matrix.
  $(\vec r,\vec r')$ runs over all possible nearest-neighbor pairs on the square lattice.

 The 2D $q$-state quantum Potts model defined in Eq.~(\ref{ham1})
 is invariant with respect to the $q$-way unitary transformations, i.e.,
\begin{equation}
 U_m : \left\{
        \begin{array}{ccc}
        M^{[\vec r]}_{x,p} & \rightarrow & \big(\omega^{p}_q\big)^{m-1} M^{[\vec r]}_{x,p}
        \\
         M^{[\vec r]}_z & \rightarrow & M^{[\vec r]}_z
 \end{array} \right. ,
 \label{transformation}
 \end{equation}
 where  $\omega_q =\exp[i\theta]$ with characteristic angle $\theta=2\pi/q$ and $m \in [1, q]$.
 These unitary transformations, of the form
 $U_m H_q U^\dagger_m = H_q$,
 imply that the 2D $q$-state Potts model possesses a $Z_q$ symmetry.
 According to the spontaneous symmetry breaking mechanism,
 for the $Z_q$ broken-symmetry phase,
 the system has a $q$-fold degenerate groundstate.
 The $Z_q$ broken-symmetry phase
 can be characterized by the nonzero value of a local order.
 If $\lambda \gg 1$, Eq.~(\ref{ham1}) becomes $H_q \approx -\sum_{\vec r} M^{[\vec r]}_{z}$
 and then the transformation in Eq.~(\ref{transformation})
 is nothing but the identity transformation, i.e., $U_m = I_q$.
 The groundstate is non-degenerate in the $Z_{q}$
 symmetry phase.

 \section{$\mbox{\bf i}$PEPS algorithm}

 To demonstrate numerically
 detecting the $q$-fold degenerate groundstate
 in the 2D $q$-state quantum Potts model,
 we employ the infinite projected entangled-pair state (iPEPS) algorithm~\cite{iPEPS,iTEBDorus,xiang}.
 Let us then briefly explain the iPEPS algorithm as follows.
 Consider an infinite
 2D square lattice where each site is labeled by a vector
 $\vec{r} = (x,y)$.
 Each lattice site can be represented by a local Hilbert space
 $V^{[\vec{r}]} \cong {C}^d$ of finite dimension $d$.
 The Hamiltonian $H_q$ with the nearest neighbor interactions on the square lattice
 is invariant under shifts by one lattice site.
 $H_q= \sum_{(\vec{r},\vec{r}')} h_q^{[\vec{r},\vec{r}']}$
 can decompose as a sum
 of terms $ h_q^{[\vec{r},\vec{r}']}$ involving pairs of nearest neighbor sites.
 In the infinite 2D square lattice,
 the state $|\Psi\rangle$ can be constructed in terms of only two tensors $A^{[x,x+2y]}$ and $B^{[x,x+2y+1]}$
 with $x, y \in \mathbb{Z}$,
 which the state $|\Psi\rangle$ is invariant under shifts by two lattice sites.
%
%
 The five index tensors $A^{[\vec{r}]}_{sudlr}$ and $B^{[\vec{r'}]}_{sudlr}$ are made up of
 complex numbers labeled by one physical index $s$ and the four
 inner indices $u$, $d$, $l$ and $r$.
 The physical index $s$
 runs over a basis of $V^{[\vec{r}]}$ so that $s=1,\ldots,d$.
 Each inner index takes $D$ values as a bond
 dimension and connects a tensor with its nearest
 neighbor tensors.
 In the iPEPS representation, thus,
 one can prepare a random initial state $|\Psi(0)\rangle$ numerically.

 To calculate a groundstate of the system, the idea is to use the infinite time-evolving block decimation
 (iTEBD) algorithm, i.e., the imaginary time evolution of the prepared initial state $|\Psi(0)\rangle$
 driven by the Hamiltonian $H_q$, i.e.,
   $ |\Psi(\tau)\rangle =
 {e^{-H_q\tau} |\Psi(0)\rangle}/||e^{-H_q\tau}
 |\Psi(0)\rangle||$~\cite{iPEPS}.
 Using a Suzuki-Trotter expansion of the time-evolution operator $U=e^{-H_q \tau}$ \cite{suzuki},
 and then updating the tensors as $A'^{[\vec{r}]}_{sudlr}$ and $B'^{[\vec{r'}]}_{sudlr}$  after
 applying each of these extended operators
 leads to an iPEPS groundstate of the system $H_q$ for a large enough $\tau$.
 For a time slice, the evolution procedure
 has a contraction process in order to get the effective environment for a pair of the tensors
 $A$ and $B$~\cite{iPEPS,iTEBDorus}.
 Practically, a sweep technique ~\cite{TI_MPS},
 originally devised for an MPS algorithm applied to one-dimensional quantum
 systems with periodic boundary conditions ~\cite{PBC_MPS}, can be used to compute two updated tensors
 $A'$ and $B'$.
 After the time-slice evolution, then, all the tensors are updated.
 This procedure is repeatedly performed until the system energy converges to a ground-state
 energy that yields a groundstate wave function in the iPEPS
 representation.

\section{degenerate groundstates and quantum fidelity}

Once one obtains an iPEPS ground state $|\psi^{(n)} \rangle$
with the $n$th randomly chosen initial state, one can define a quantum fidelity
 $F(|\psi^{(n)}\rangle, |\phi\rangle) = |\langle \psi^{(n)}|\phi\rangle|$  between the groundstate and
 a chosen reference state $|\phi\rangle$.
 Actually, $F(|\psi^{(n)}\rangle, |\phi\rangle)$ means
 a projection of $|\psi^{(n)}\rangle$ onto $|\phi\rangle$.
 If the system has only one ground state for the parameters,
 the projection value $F(|\psi^{(n)}\rangle, |\phi\rangle)$
 has only one constant value with the random initial states.
 If $F(|\Psi^{(n)}\rangle, |\phi\rangle)$ has $n$ projection values
 with the random initial states, the system must have $n$ degenerate
 ground states for the fixed system parameters.
 With different initial states for a fixed system parameter,
 one can then determine how many ground states exist from how many projection values exist.

 For our numerical calculation, we choose the numerical reference state $|\phi\rangle$ randomly.
 The quantum fidelity $F(|\psi^{(n)}\rangle,|\phi\rangle)$ asymptotically scales
 as $F(|\psi^{(n)}\rangle,|\phi\rangle) \sim d^{L}$,
 where $L=L_x \times L_y$ is the size of the two-dimensional square lattice.
 In Ref. \cite{zhou1,zhouzhlb,zhwlzh,dhzhzhou,Su,Zhou08}, the fidelity per lattice site (FLS) is defined as
\begin{equation}
  \ln d(|\psi^{(n)}\rangle,|\phi\rangle)
  = \lim_{L \rightarrow \infty}
      \frac{1}{L} \ln F(|\psi^{(n)}\rangle,|\phi\rangle).
  \label{fidelity}
\end{equation}
 The FLS is well defined in the
 thermodynamic limit, even if $F$ becomes trivially zero.
 The FLS is within the range $0 \leq d(|\psi^{(n)}\rangle,|\phi\rangle) \leq 1$.
 If $|\psi^{(n)}\rangle=|\phi\rangle$ then $d = 1$.
 Within the iPEPS approach, the FLS is given by the largest
 eigenvalue of the transfer matrix \cite{Zhou08}.
 In this section, we will demonstrate explicitly how to
 detect degenerate groundstates of the 2D $q$-state quantum Potts model
 by means of the quantum fidelity in Eq.
 (\ref{fidelity}).

\begin{figure}
\includegraphics[width=0.3\textwidth]{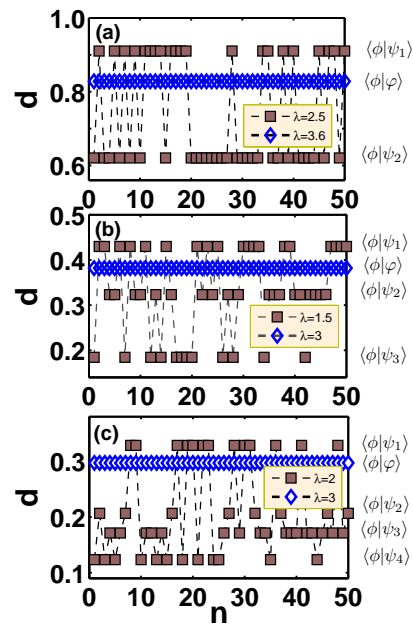}
 \caption{(color online)
 Groundstate quantum fidelity per site $d$ for (a) the quantum Ising
 model, (b) the quantum three-state Potts model and (c) the
 quantum four-state Potts model on the square lattice
 with an arbitrary reference state as a function of random
 initial state trials $n$.
 Here, an arbitrary reference state $|\phi\rangle$
 is chosen numerically.
 The numerical iPEPS groundstates $|\psi\rangle$
 are in the broken symmetry phase with transverse coupling
 (a) $\lambda=2.5$, (b) $\lambda= 1.5$ and (c) $\lambda=2$.
 The state $|\varphi\rangle$ is
 in the symmetry phase with  (a) $\lambda=3.6$, (b) $\lambda=3.0$ and (c) $\lambda=3.0$.
 Note that (a) two, (b) three and (c) four different values of the fidelity indicate
 that there are (a) two, (b) three and (c) four degenerate ground states in the symmetry broken phase.}
   \label{fig1}
\end{figure}

To begin, we choose $\lambda = 2.5$ and $3.6$ for the quantum Ising model ($q=2$),
 $\lambda = 1.5$ and $3.0$ for the three-state quantum Potts model ($q=3$), and
 $\lambda = 2.0$ and $3.0$ for the four-state quantum Potts model ($q=4$).
 For each given $\lambda$, the iPEPS goundstates are calculated with fifty different
 randomly chosen initial states, i.e., $n=50$.
 To calculate the FLS $d$,
 the arbitrary numerical reference state $| \phi \rangle$ is also chosen randomly.
 In Fig.~\ref{fig1}, we plot the FLS $d$ as a function of the random initial
 states for (a) $q=2$, (b) $q=3$ and (c) $q=4$.
 For the Ising model, Fig.~\ref{fig1}(a) shows that there are two different values of the FLS for $\lambda=2.5$,
 while there exists only one value of the FLS for $\lambda=3.6$.
 This implies that, for $\lambda = 2.5$, the Ising model system is in the $Z_2$ broken-symmetry phase,
 while the system is in the symmetry phase for $\lambda = 3.6$.
 We label the two degenerate groundstates by $|\psi_1\rangle$ and  $|\psi_2\rangle$ for each value of
 the FLS for $\lambda=2.5$.
 For $\lambda=3.6$ the groundstate is denoted by $|\varphi\rangle$.

For the three-state Potts model, Fig.~\ref{fig1}(b)
 shows that there are three different values of the FLS for $\lambda=1.5$,
 while there is only one value of the FLS for $\lambda=3.0$.
 Thus, for $\lambda = 1.5$, the three-state Potts model is in the $Z_3$ broken-symmetry phase,
 while the system is in the symmetry phase for $\lambda = 3.0$.
 We label the three degenerate groundstates from each value of the FLS for $\lambda=1.5$ by
 $|\psi_1\rangle$, $|\psi_2\rangle$ and  $|\psi_3\rangle$.
 For $\lambda=3.0$, the groundstate is denoted
 by $|\varphi\rangle$ in the symmetry phase.

 Consistently, one may expect that
 there are four degenerate groundstates
 in the $Z_4$ broken-symmetry phase
 while there exists only one groundstate in the symmetry phase.
 Indeed, for $q=4$, Fig.~\ref{fig1}(c) shows the four degenerate groundstates for
 $\lambda = 2.0$ and the one groundstate for $\lambda=3.0$.
 Athough we have demonstrated how to detect all of the degenerate groundstates
 only for the cases $q=2$, $q=3$ and $q=4$ in this study,
 one may detect $q$-fold degenerate groundstates in
 the 2D $q$-state quantum Potts model on the infinite square lattice for any $q$.
 Also, the above results imply
 that the phase transition points $\lambda_c$
 should exist between (a)
 $\lambda = 2.5$ and $\lambda = 3.6$ for the Ising model,
 (b) $\lambda = 1.5$ and $\lambda = 3.0$ for the three-state Potts model,
 and (c) $\lambda=2.0$ and $\lambda=3.0$ for the four-state Potts model.
 The nature of the phase transitions will be discussed in the next section.

 In order to ensure that we detect all degenerate groundstates,
 we have chosen over fifty random initial states for each $q$.
 The probability $P_q(n)$ that the system is in each groundstate
 for the broken-symmetry phase is shown to be
 $P_2(n) \simeq 1/2$ (Ising model), $P_3(n) \simeq 1/3$ (three-state Potts model)
 and $P_4(n) \simeq 1/4$ (four-state Potts model)
 in the broken-symmetry phase.
 For given $q$, then,
 with a large number of random initial state trials,
 one may detect the $q$ degenerate iPEPS groundstates
 with the probability $P_q(n \rightarrow \infty)=1/q$ for finding each degenerate groundstate
 in the $Z_q$ broken-symmetry phase.
 Consequently, in the 2D $q$-state quantum Potts model on the infinite square lattice,
 it is shown that all of the $q$-fold degenerate states for the $Z_q$ broken-symmetry phase can be detected
 by using the quantum fidelity with an arbitrary reference state in Eq. (\ref{fidelity}).

\section{Multiple-bifurcations of the FLS and phase transitions}

 In the Landau-Ginzburg-Wilson paradigm for quantum phase transitions, as is well-known,
 spontaneous symmetry breaking leads to a system having degenerate groundstates
 in its broken-symmetry phase.
 This means that
 the degenerate groundstates in the broken-symmetry phase
 exist until the system reaches its phase boundaries, i.e., its phase transition point.
 In the $q$-state quantum Potts model,
 the $q$-fold degenerate groundstates for the broken-symmetry phases
 become one groundstate at phase transition points.
From the perspective of quantum fidelity,
 the $q$ different values of the FLS, which indicate
 the $q$ different degenerate groundstates,
 should collapse into one value of the FLS at a phase transition point.

\begin{figure}
\includegraphics[width=0.3\textwidth]{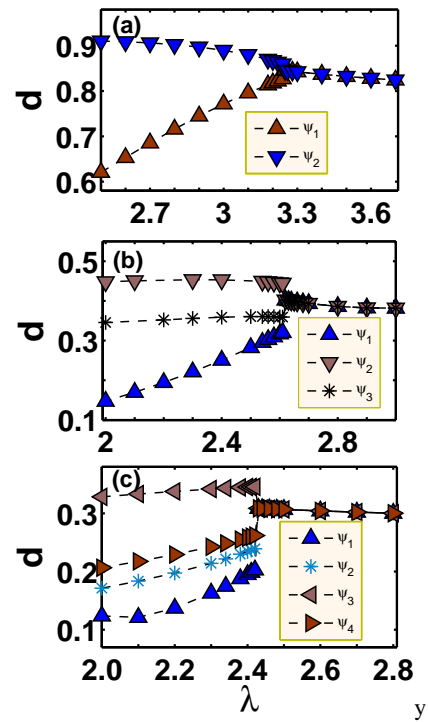}y
\caption{(color online)
  Groundstate quantum fidelity per site $d$ for (a) the quantum Ising model,
 (b) the quantum three-state Potts model and (c) the quantum four-state Potts model
 as a function
 of the transverse magnetic field $\lambda$ with the truncation dimension (a) $\chi=4$,
 (b) $\chi=6$ and (c) $\chi=4$.  In the broken symmetry phase, the $q$ branches of the FLS
 correspond to the $q$ degenerate groundstates.  As the magnetic field crosses
 the critical point $\lambda_{c}$, the FLS shows multiple bifurcations with (a) two, (b) three and
 (c) four branches in the broken symmetry phase.}
\label{fig2}
\end{figure}

 In order to see such expected behavior of the FLS,
 we have detected the iPEPS degenerate groundstates by varying
 the transverse magnetic field $\lambda$.
 From the detected iPEPS degenerate groundstates,
 we plot the FLS as a function of
 $\lambda$ for $q=2$, $3$ and $4$ in Fig.~\ref{fig2}.
 Figure~\ref{fig2} shows clearly that, as the transverse magnetic field decreases,
 the single value of the FLS in the broken-symmetry phase
 branches into $q$ values.
 The branch points of the FLS are estimated numerically to be
 $\lambda=3.23$ for $q=2$, $\lambda =2.616$ for $q=3$
 and $\lambda=2.43$ for $q=4$ (cf Fig.~\ref{fig2}).
 In fact, the branch points are expected to be the phase transition points obtained from
 the local order parameters, which will be shown in the next section.
 Such branching behavior of the FLS can be called {\it multiple
 bifurcation} and such a branch point a {\it multiple bifurcation point}.

 Moreover, it should be noted that for $q=2$ (the
 Ising model) the branching is continuous, while for $q=3$ and $q=4$
 the branching is abrupt.
 Such continuous (discontinuous) behavior of the FLS
 indicates a continuous (discontinuous) phase transition.
 As a result,
 the FLS in Eq. (\ref{fidelity}) can distinguish between continuous and discontinuous quantum
 phase transitions.
 In this way the $q$-state quantum Potts model on the square lattice undergoes
 a continuous (discontinuous) phase transition
 for $q \leq2$ ($q >2$).

 As already mentioned,
 we have chosen several reference states for the quantum fidelity calculation.
 Any randomly chosen reference state except for the system groundstates
 gives the same number of groundstates and the same critical point,
 though the amplitudes of the quantum fidelity depend on a chosen reference state.
 Consequently, it has been demonstrated that
 the quantum fidelity between degenerate groundstates
 and an arbitrary reference state
 can detect a critical point.
 However, it  should be stressed that our emphasis here is not in obtaining accurate estimates
 for the critical points $\lambda_c$.
 Rather our emphasis is on the general framework for detecting degenerate
 groundstates in a 2D quantum system using quantum fidelity to determine continuous or discontinuous phase transitions
 due to a spontaneous symmetry breaking.
 Indeed, the estimate $\lambda_c \sim 3.23$ obtained for the critical point of the quantum
 transverse Ising model on the square lattice compares rather poorly with the most accurate
 current estimate $\lambda_c = 3.044$, obtained using quantum Monte Carlo \cite{blote}.
 For this model, previous studies using iPEPS yield estimates of 3.06 \cite{iPEPS} and 3.04 \cite{ctmrg},
 where the latter estimate involves a modification using the corner transfer matrix renormalization group.
 For the three-state quantum Potts model on the square lattice, the estimate $\lambda_c \sim 2.616$
 is closer to the known estimate $\lambda_c \sim 2.58$ \cite{hamer_q=3}.
 As far as we are aware, there are no other estimates to compare with our result $\lambda_c \sim 2.43$
 for the four-state quantum Potts model on the square lattice.
 We note that in each case our estimates could be improved by using a more refined updating scheme in the iPEPS
 algorithm, rather than the simplified updating scheme used.


\begin{figure}
 \includegraphics[width =0.48\textwidth]{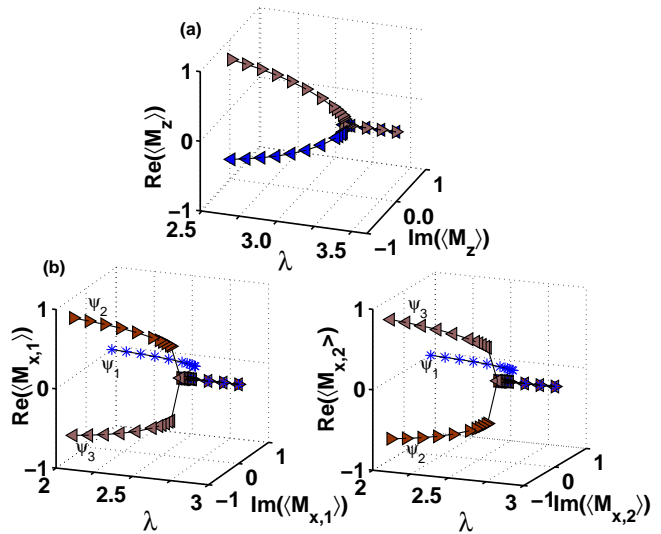}
  \caption{(color online)
Magnetization (a) $\langle M_{z}\rangle$ as a function of the transverse magnetic field
$\lambda$ for the quantum Ising model obtained with truncation dimension $\chi=4$.
The critical point is estimated to be at $\lambda_c=3.23$.
Magnetization (b) (left panel) $M_{x,1}$
and (b) (right panel) $M_{x,2}$ as a function of the transverse magnetic field $\lambda$ for the quantum
 three-state Potts model obtained with truncation dimension $\chi=6$.
 The critical point is estimated to be at $\lambda_c=2.616$.}
 \label{fig3}
\end{figure}

 \section{Order parameters}

 According to the Landau theory of spontaneous symmetry breaking,
 a broken-symmetry phase is characterized by nonzero
 values of a local observable -- the local order parameter.
 As discussed \cite{Su}, spontaneous symmetry breaking leads to
 a degenerate groundstate for the broken-symmetry phase.
 Consequently, the relations between the local order parameters calculated from degenerate groundstates
 are determined by a symmetry group of the system Hamiltonian.
 Here we will show a relation between the local order parameters within
 the subgroup of the symmetry group of the system Hamiltonian.

 Let us first discuss the local magnetization for the quantum Ising model.
 In Fig.~\ref{fig3}(a), we plot the magnetization
 $\langle M_{x}\rangle_{m}$ as a function of the traverse
 magnetic field $\lambda$.
 The magnetizations disappear gradually to zero at the numerical critical
 point $\lambda_c$.
 For the broken-symmetry phase $\lambda < \lambda_c$,
 the magnetization is calculated from each of the two degenerate
 groundstates, where each groundstate wavefunction is denoted by $|\psi_m\rangle$, with $m \in \{1,2\}$.
 The magnetizations are related to each other by $\langle M_{x}\rangle_{1}=-\langle M_{x}\rangle_{2}$.
 Then, for a given magnetic field, the relation between the two magnetizations in the complex
 magnetization plane can be regarded as a rotation
 characterized by the value $\omega_2 = \exp[2\pi i/2]$, i.e.,
 $\langle M_{x}\rangle_{1}=\omega^{-1}_2 \langle M_{x}\rangle_{2}$.
 The two degenerate groundstates give the same value for the
 $z$-component magnetizations, i.e., $\langle M_{z}\rangle_1=\langle M_{z}\rangle_2$.
 This implies that the Ising model Hamiltonian
 is invariant under the unitary transformations
 $U_1=I$ and $U_2$ in Eq. (\ref{transformation}),
 but (as expected) the two degenerate groundsates
 are  not invariant under the unitary transformation $U_2$ in the $Z_2$ broken-symmetry phase.
 Thus, the characteristic rotation angles between the different magnetizations
 are $\theta=0$ and $\theta = \pi$.
 The relation between the magnetizations can be written as
 $\langle M_{x}\rangle_m=g_2 \langle M_{x}\rangle_{m'}$ with
 $g_2 \in \{ I, \omega_2 \}$.
 Further, the magnetizations are shown to exhibit a bifurcation behavior, similar to the FLS.
 The continuous behavior of the two magnetizations also shows that the phase transition
 is continuous.

 For the three-state quantum Potts model,
 in Fig.~\ref{fig3}(b), we display the magnetizations $\langle M_{x,1}\rangle$
 and $\langle M_{x,2}\rangle$ as a function of the traverse magnetic field.
 Note that all of the absolute values of the magnetizations
 $\langle \psi_m | M_{x,p} | \psi_m \rangle \equiv \langle M_{x,p}\rangle_m$
 are the same at a given magnetic field $\lambda$.
 Here, we have chosen the state $|\psi_1\rangle$ that gives
 a real value of the magnetization, i.e.,
 $\langle M_{x,1}\rangle_1$ and $\langle M_{x,2}\rangle_1$ are real.
 In contrast to the quantum Ising model,
 all of the magnetizations disappear abruptly to zero at the critical
 point $\lambda_c$, which indicates that
 the phase transition is a discontinuous.
For the broken-symmetry phase $\lambda < \lambda_c$,
the magnetization is calculated from each the three degenerate
groundstates denoted by $|\psi_m\rangle$.
 For a given magnetic field, the magnetizations in the complex
 magnetization plane are related with a characteristic rotation
 $\omega_3 = \exp[2\pi i/3]$.
 Actually in Fig.~\ref{fig3}(b) it is observed that
 the magnetizations are related to one another by
 $\langle M_{x,1}\rangle_1=\omega^{-1}_3 \langle M_{x,1}\rangle_2
 =\omega^{-2}_3 \langle M_{x,1}\rangle_3$
 and
 $\langle M_{x,2}\rangle_1=\omega^{-2}_3 \langle M_{x,2}\rangle_2
 =\omega^{-4}_3 \langle M_{x,2}\rangle_3$.
 Each groundstate wavefunction gives the relations
 $\langle M_{x,1}\rangle_1=\langle M_{x,2}\rangle_1$,
 $\langle M_{x,1}\rangle_2=\omega^{-1}_3 \langle M_{x,2}\rangle_2$ and
 $\langle M_{x,1}\rangle_3=\omega^{-2}_3 \langle M_{x,2}\rangle_3$.
 The three degenerate groundstates also give the same value for the
 $z$-component magnetizations, i.e., $\langle M_{z}\rangle_1=\langle M_{z}\rangle_2
 =\langle M_{z}\rangle_3$.
 These imply that the three-state quantum Potts model Hamiltonian
 is invariant under the unitary transformations
 $U_1=I$, $U_2$ and $U_3$ with $\omega_3 = \exp[2\pi i/3]$ in Eq. (\ref{transformation}),
 but the three degenerate groundsates
 are not invariant under the unitary transformations $U_2$ and $U_3$ in the $Z_3$ broken-symmetry phase.
 Thus the characteristic magnetization rotation angles
 are $\theta = 0$, $2\pi/3$ and $4\pi/3$.
 The magnetizations obey
 the relations $\langle M_{x,p}\rangle_m=g_3 \langle M_{x,p'}\rangle_{m'}$ with
 $g_3 \in \{ I, \omega_3, \omega^2_3 \}$.

 Based on the above relations between the magnetizations for $q=2$ and $q=3$,
 we can infer a general relation between the magnetizations for the
 $q$-state quantum Potts model on the square lattice.
 For any $q$, the relations are given by
\begin{subequations}
   \label{eq4}
\begin{eqnarray}
 \langle M_{x,p} \rangle_m
   &=& \omega^{p'(1-m')-p(1-m)}_q\langle M_{x,p'} \rangle_{m'}, \\
    \langle M_{z} \rangle_m
   &=& \langle M_{z} \rangle_{m'}.
\end{eqnarray}
\end{subequations}
 The magnetizations $M_{x,p}$ with respect to a different degenerate groundstate
 (i.e., $p=p'$)
 satisfy $\langle M_{x,p} \rangle_m = \omega^{p(m-m')}_q \langle M_{x,p}
 \rangle_{m'}$ as deduced from Eq.~(\ref{eq4}).
 For one of the degenerate groundstate wavefunction (i.e., $m=m'$),
 the magnetizations of the operators $M_{x,1},\ldots, M_{x,q-2}$ and
 $M_{x,q-1}$ satisfy
 $\langle M_{x,p} \rangle_m = \omega^{(p'-p)(1-m)}_q \langle M_{x,p'}
 \rangle_m $ as deduced from Eq.~(\ref{eq4}).
 These results show that, in the complex magnetization plane,
 the rotations between the magnetizations
 for a given magnetic field
 are determined by the characteristic rotation angles $\theta = 0, 2\pi/q,
 4\pi/q, \ldots, 2(q-1)\pi/q$.
 As a result, the relations between the order parameters in Eq.~(\ref{eq4}a) can be rewritten as
\begin{subequations}
\begin{eqnarray}
 \langle M_{x,p}\rangle_m &=& g_q \langle M_{x,p'}\rangle_{m'} ,
 \\
 g_q
  &\in& \{ I, \omega_q, \omega^2_q, \cdots, \omega^{q-1}_q \}.
\end{eqnarray}
 \label{eq5}
\end{subequations}
Equation~(\ref{eq5}) shows clearly that
 the 2D $q$-state quantum Potts model on the square lattice has
 the discrete symmetry group $Z_q$ consisting of $q$ elements.


\begin{figure}
 \includegraphics[width =0.45\textwidth]{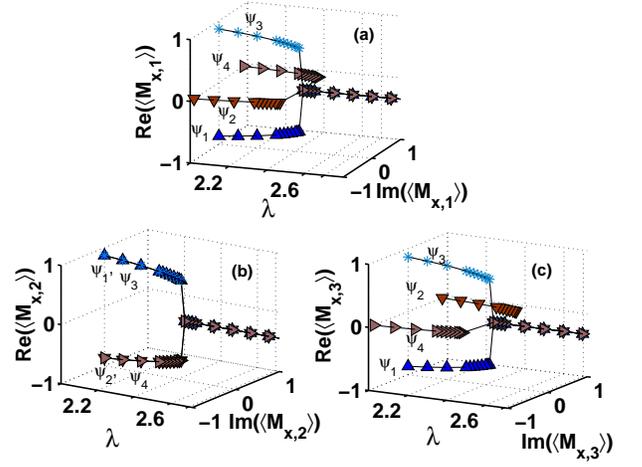}
  \caption{(color online)
      Magnetization (a) $\langle M_{x,1} \rangle$, (b) $\langle M_{x,2} \rangle$
      and (c) $\langle M_{x,3} \rangle$
      as a function of the transverse magnetic field $\lambda$ for the quantum four-state Potts model obtained
      with truncation dimension $\chi=4$.  For the broken symmetry phase the magnetizations follow from
      each of the four degenerate groundstates.
 The critical point is estimated to be at $\lambda_c=2.43$.
}

 \label{fig4}
\end{figure}

 One can show that the characteristic relations
 between the magnetizations in Eqs. (\ref{eq4}) and (\ref{eq5}) hold for
 the four-state quantum Potts model.
 In Fig.~\ref{fig4} we plot the magnetizations $\langle M_{x,1}\rangle$,
 $\langle M_{x,2}\rangle$ and $\langle M_{x,3}\rangle$
 as a function of the traverse magnetic field $\lambda$.
 The magnetizations indicate that the phase transition
 is discontinuous.
 Also, the absolute values of the magnetizations
 have the same values at a given magnetic field
 and the magnetizations in the complex
 magnetization plane have a relation between them under rotation
 characterized by the value $\omega_4 = \exp[2\pi i/4]$.
 The degenerate groundstates give the same values
 $\langle M_{z}\rangle_1=\langle M_{z}\rangle_2
  =\langle M_{z}\rangle_3=\langle M_{z}\rangle_4$
 for the $z$-component magnetizations.

 In Fig.~\ref{fig4}, we also observe that
 for a given magnetic field $\lambda < \lambda_c$,
 the relations between the magnetizations are
 $\langle M_{x,1}\rangle_1=\omega^{-1}_4 \langle M_{x,1}\rangle_2
 =\omega^{-2}_4 \langle M_{x,1}\rangle_3
 =\omega^{-3}_4 \langle M_{x,1}\rangle_4$ from Fig.~\ref{fig4}(a),
 $\langle M_{x,2}\rangle_1=\omega^{-2}_4 \langle M_{x,2}\rangle_2
 =\omega^{-4}_4 \langle M_{x,2}\rangle_3
 =\omega^{-6}_4 \langle M_{x,2}\rangle_4$ from Fig.~\ref{fig4}(b)
 and
 $\langle M_{x,3}\rangle_1=\omega^{-3}_4 \langle M_{x,3}\rangle_2
 =\omega^{-6}_4 \langle M_{x,3}\rangle_3
 =\omega^{-9}_4 \langle M_{x,3}\rangle_4$ from Fig.~\ref{fig4}(c).
 Also, for each groundstate wavefunction the magnetizations
 obey the relations
 $\langle M_{x,1}\rangle_1=\langle M_{x,2}\rangle_1=\langle M_{x,3}\rangle_1$,
 $\langle M_{x,1}\rangle_2=\omega^{-1}_4 \langle M_{x,2}\rangle_2
 =\omega^{-2}_4 \langle M_{x,3}\rangle_2$,
 $\langle M_{x,1}\rangle_3=\omega^{-2}_4 \langle M_{x,2}\rangle_3
 =\omega^{-4}_4 \langle M_{x,3}\rangle_3$,
and
 $\langle M_{x,1}\rangle_4=\omega^{-3}_4 \langle M_{x,2}\rangle_4
 =\omega^{-6}_4 \langle M_{x,3}\rangle_4$.
As expected from Eqs. (\ref{eq4}) and (\ref{eq5}),
 these results show that in the complex magnetization plane
 the rotations between the magnetizations
 for a given magnetic field
 are determined by the characteristic rotation angles $\theta = 0$, $2\pi/4$,
 $4\pi/4$, and $6\pi/4$, i.e.,
 $\langle M_{x,p}\rangle_m=g_4 \langle M_{x,p'}\rangle_{m'}$ with
 $g_4 \in \{ I, \omega_4, \omega^2_4, \omega^3_4 \}$.

 The general results in Eqs.~(\ref{eq4}) and (\ref{eq5}) hold for any $q$
 in the 2D $q$-state quantum Potts model on the square lattice.
 It is shown how each order parameter transforms under
 the subgroup for the $Z_q$ symmetry group
 in the 2D $q$-state quantum Potts model
 on the infinite square lattice within the spontaneous symmetry mechanism.

%

\section{Summary}

 We have investigated the quantum fidelity in
 the two-dimensional $q$-state quantum Potts model by employing the iPEPS algorithm
 on the infinite square lattice.
The degenerate iPEPS groundstates have been successfully detected using
the quantum fidelity.
 We have shown
 (i)
 that each of the degenerate groundstates possesses its own order described
 by a corresponding order parameter -- the magnetization $\langle M_{x,p}\rangle_{m}$ --
 in the broken-symmetry phases,
 (ii)
 that each order parameter, which is nonzero only in the broken-symmetry phases,
 distinguishes the ordered phase from
 the disordered phases, which results in the multiple bifurcation of the order parameters
 at the phase transition points, and
 (iii)
 further, how each order parameter transforms under
 the subgroup of the $Z_q$ symmetry group.

 In line with previous studies, we found that
 the $q$-state quantum Potts model on the square lattice
 undergoes a discontinuous (first-order) phase transition for $q=3$ and $q=4$, and
 a continuous phase transition for the quantum Ising model ($q=2$).
 Consequently, we have demonstrated that
 (i)
 the multiple bifurcations of the quantum fidelity
 result from the spontaneous $Z_q$-symmetry breaking in the broken-symmetry phase,
 (ii)
 that the multiple bifurcation points of the quantum fidelity, corresponding to the multiple bifurcation
 of the order parameters, correspond to the phase transition points, and
 (iii)
 the (dis-)continuous behavior of the quantum fidelity indicates
 that the system undergoes a (dis-)continuous quantum phase transition at the multiple bifurcation
 points.

Our results show conclusively that the quantum fidelity can be used for detecting degenerate groundstates and
 phase transition points, and for determining continuous or discontinuous phase transitions
 due to a spontaneous symmetry breaking, without knowing any details of a broken symmetry
 between a broken-symmetry phase and a symmetry phase as a system
 parameter crosses its critical value (i.e., at a multiple bifurcation point).

\acknowledgements

  Y.W.D. acknowledges support from the Fundamental Research Funds for the Central Universities
  (Project No. CDJXS11102214) and the Chongqing University Postgraduate's Science and Innovation
  Fund (Project No. 200911C1A0060322).
  This work was supported by the
  National Natural Science Foundation of China (Grant No. 11374379 and Grant No. 11174375).
  M.T.B.  acknowledges support from the 1000 Talents Program of China.

 
\end{document}